\documentclass[aps,prb,superscriptaddress,amsfonts,amsmath,amssymb,twocolumn,showpacs,floatfix]{revtex4-1}
\usepackage{url}
\usepackage{bm}
\usepackage{graphicx}
\usepackage{amsmath}
\usepackage{amstext}
\usepackage{amssymb}
\usepackage{amsfonts}
\usepackage{amsbsy}
\usepackage{verbatim}
\usepackage{color}
\usepackage[colorlinks=true, urlcolor=blue, linkcolor=blue, citecolor=blue, pdftex]{hyperref}
\usepackage{multirow}
\usepackage{pdfpages}

\begin{document}

\title{Projected wave function study of ${\bf \mathbb{Z}_{2}}$ spin liquids on the 
kagome lattice for the spin-$\frac{1}{2}$ quantum Heisenberg antiferromagnet}

\author{\href{http://www.lpt.ups-tlse.fr/spip.php?article38&lang=en}{Yasir Iqbal}}
\email[]{yasir.iqbal@irsamc.ups-tlse.fr}
\affiliation{Laboratoire de Physique Th\'eorique UMR-5152, CNRS and 
Universit\'e de Toulouse, F-31062 France}
\author{\href{http://www.democritos.it/curri/federico.becca.php}{Federico Becca}}
\email[]{becca@sissa.it}
\affiliation{Democritos National Simulation Center, Istituto
Officina dei Materiali del CNR and Scuola Internazionale Superiore di Studi 
Avanzati (SISSA), Via Bonomea 265, I-34136 Trieste, Italy}
\author{\href{http://www.lpt.ups-tlse.fr/spip.php?article32}{Didier Poilblanc}}
\email[]{didier.poilblanc@irsamc.ups-tlse.fr}
\affiliation{Laboratoire de Physique Th\'eorique UMR-5152, CNRS and 
Universit\'e de Toulouse, F-31062 France}

\date{\today}

\begin{abstract}
Motivated by recent density-matrix renormalization group (DMRG) calculations
[Yan, Huse, and White, 
\href{http://dx.doi.org/10.1126/science.1201080}{Science {\bf 332}, 1173 (2011)}],
which claimed that the ground state of the nearest-neighbor spin-$1/2$ 
Heisenberg antiferromagnet on the kagome lattice geometry is a fully gapped 
spin liquid with numerical signatures of $\mathbb{Z}_{2}$ gauge structure, 
and a further theoretical work [Lu, Ran, and Lee, 
\href{http://dx.doi.org/10.1103/PhysRevB.83.224413}{Phys. Rev. B {\bf 83}, 224413 (2011)}], which gave a  
classification of all Schwinger-fermion mean-field fully symmetric 
$\mathbb{Z}_{2}$ spin liquids on the kagome lattice, we have thoroughly studied 
Gutzwiller-projected fermionic wave functions by using quantum variational 
Monte Carlo techniques, hence implementing exactly the constraint of one 
fermion per site. In particular, we investigated the energetics of all 
$\mathbb{Z}_{2}$ candidates (gapped and gapless) that lie in the neighborhood 
of the energetically competitive U($1$) gapless spin liquids. 
By using a state-of-the-art optimization method, we were able to conclusively 
show that the U($1$) Dirac state is remarkably stable with respect to all 
$\mathbb{Z}_{2}$ spin liquids in its neighborhood, and in particular for 
opening a gap toward the so-called $\mathbb{Z}_{2}[0,\pi]\beta$ state, 
which was conjectured to describe the ground state obtained by the DMRG method.
Finally, we also considered the addition of a small second nearest-neighbor exchange 
coupling of both antiferromagnetic and ferromagnetic type, and obtained similar
results, namely, a U($1$) Dirac spin-liquid ground state. 
\end{abstract}
\maketitle

{\it Introduction}.
The nearest-neighbor (NN) spin-$1/2$ quantum Heisenberg antiferromagnet 
(QHAF) on the kagome lattice provides ideal conditions for the amplification
of quantum fluctuations and a consequent stabilization of an exotic 
magnetically disordered ground state, which may be a valence-bond crystal 
(VBC)~\cite{Marston-1991,Hastings-2000,Nikolic-2003,Singh-2007,DP-2010} 
or a spin liquid (SL) with fractionalized 
excitations.~\cite{Anderson-1973,Anderson-1987,White-2010}
Recent experiments have unanimously pointed toward a SL behavior;~\cite{Mendels-2008,Mendels-2007,Bert-2007,Nocera-2008,Nocera-2006,Shores-2007,Huang-2007,Kamenev-2008}
in particular, Raman spectroscopic data on a nearly perfect spin-$1/2$ kagome
compound with Heisenberg couplings (the so-called Herbertsmithite) suggested
a gapless (algebraic) SL.~\cite{Wulferding-2008} On the theoretical side, 
the question is still wide open and intensely debated. On the one hand, 
series expansion provided evidence that a VBC with a 36-site unit cell has 
lower energy than other proposed competing states.~\cite{Singh-2007}
On the other hand, it was shown that, within the class of Gutzwiller-projected 
fermionic wave functions, a particular algebraic SL, the so-called U($1$) 
Dirac state, has a competing energy.~\cite{Lee-2007} 
Its properties were studied in detail in Ref.~[\onlinecite{Hermele-2008}] and 
it was argued that it can be a stable SL state. However, a recent DMRG 
study~\cite{White-2010} has challenged the above results, and proposed
that the ground state can be a fully gapped $\mathbb{Z}_{2}$ SL with 
a substantially lower energy as compared to both the above estimates. 

The $\mathbb{Z}_{2}$ SLs have the nice property that they are stable 
mean field states and can survive quantum fluctuations. Hence, they are more 
likely to occur as real physical SLs, and one can safely use the projective 
symmetry group classification of $\mathbb{Z}_{2}$ SLs beyond mean-field 
level.~\cite{Wen-2002} This complete classification of fully symmetric 
$\mathbb{Z}_{2}$ SLs on the kagome lattice was recently done in 
Ref.~[\onlinecite{Lu-2011}] within the Schwinger-fermion mean-field theory, 
resulting in an enumeration of a total of 20 $\mathbb{Z}_{2}$ mean-field 
states. Their main result was the identification of a {\it unique} gapped 
$\mathbb{Z}_{2}$ SL (called the $\mathbb{Z}_{2}[0,\pi]\beta$ state) in the 
neighborhood of the U($1$) Dirac state. Since the U($1$) Dirac SL state has 
the best variational energy among the class of U($1$) gapless SLs, in 
Ref.~[\onlinecite{Lu-2011}], it has been conjectured that the 
$\mathbb{Z}_{2}[0,\pi]\beta$ state may describe the ground state that has
been numerically observed in the DMRG study.~\cite{White-2010}

In this paper, we thoroughly investigate the possibility of any of these
$\mathbb{Z}_{2}$ SLs being stabilized as the ground state of the NN 
spin-$1/2$ QHAF, with a particular emphasis on the $\mathbb{Z}_{2}[0,\pi]\beta$
state. In practice, we compute the energy of optimized variational wave
functions that are constructed by applying the Gutzwiller projector to 
different states obtained from mean-field Hamiltonians of Schwinger fermions.
In this respect, by an exact treatment of the full projector that ensures
the one-fermion per site constraint, we go much beyond the simple mean-field
approach of Ref.~[\onlinecite{Lu-2011}]. 
We calculate the energies of all $\mathbb{Z}_{2}$ SLs which can be 
realized up to $3$rd NN in mean field {\it Ansatz} and have a non-vanishing 
$1$st NN mean-field bond. Only $12$ of the $20$ $\mathbb{Z}_{2}$ SLs
satisfy these criteria, and all of them are continuously connected to some 
U($1$) gapless SL.~\cite{Lu-2011} Our main result is that, contrary to what has
been proposed in Ref.~[\onlinecite{Lu-2011}], the $\mathbb{Z}_{2}[0,\pi]\beta$
state has a higher energy than the gapless U($1$) Dirac SL, or in 
other words the U($1$) Dirac SL is remarkably stable with respect to opening 
of a gap and consequently destabilizing into the $\mathbb{Z}_{2}[0,\pi]\beta$ 
state. We also find that all gapped $\mathbb{Z}_{2}$ SLs in the neighborhood of
another competing gapless state, the uniform resonating-valence bond (RVB) 
state, have higher energies. Moreover, we find that all $\mathbb{Z}_{2}$ SLs 
have higher energy than the gapless SL states in whose neighborhoods 
they lie.

\begin{figure}
\includegraphics[width=0.95\columnwidth]{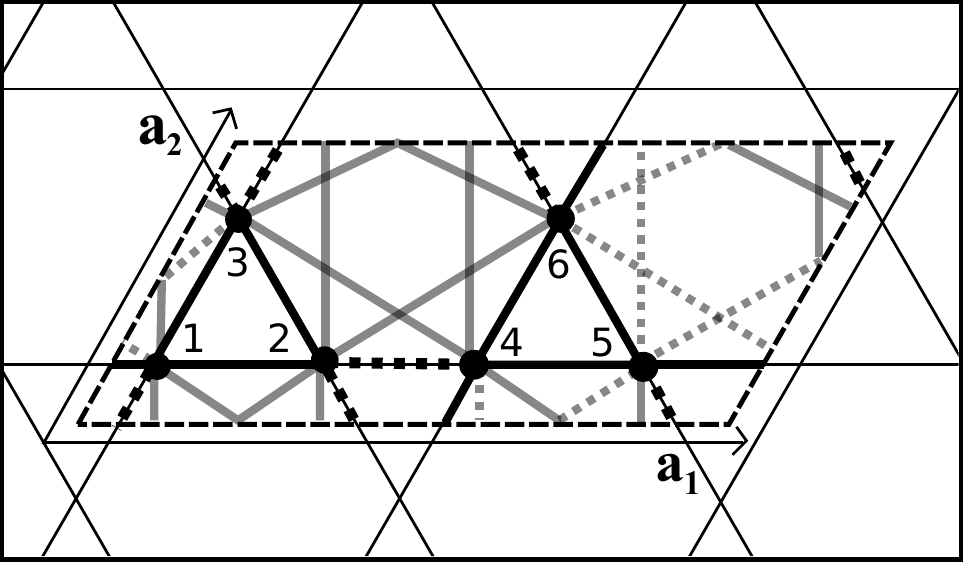}
\caption{The $\mathbb{Z}_{2}[0,\pi]\beta$ SL {\it ansatz}; black (grey) bonds 
denote $1$st NN real hopping ($2$nd NN real hopping and real spinon pairing) 
terms; solid (dashed) black bonds have ${\rm s}_{ij}=1$ ($-1$), solid (dashed)
grey bonds have $\nu_{ij}=1$ ($-1$), see Eq.~(\ref{eqn:MF-Z2}).
The $1$st NN ($2$nd NN) mean field {\it Ansatz} is written as 
${\rm U}_{\langle ij\rangle}=\pm\sigma_{3}$ 
(${\rm U}_{\boldsymbol\langle\langle ij\rangle\boldsymbol\rangle}=\pm(\chi_{2}\sigma_{3}+\Delta_{2}\sigma_{1})$). 
The SU($2$) flux ($P$), through elementary triangles (e.g., $123$) is 
$P_{123}=\sigma_{3}$, and that through triangles formed by two $1$st NN and 
one $2$nd NN bonds (e.g., $234$) is 
$P_{234}=-$($\chi_{2}\sigma_{3}+\Delta_{2}\sigma_{1}$).
Their commutator is non-zero, $[P_{123},P_{234}]=(-2i\sigma_{2})\Delta_{2}$. 
Hence, a finite $\Delta_{2}$ breaks the U($1$) gauge structure down to 
$\mathbb{Z}_{2}$, and opens up an energy gap via the Anderson-Higgs 
mechanism.~\cite{Wen-2002,Wen-1991}}
\label{fig:pic1}
\end{figure}

{\it Model and wave function}.
The Hamiltonian for the NN spin-$1/2$ Heisenberg model is
\begin{equation}
\label{eqn:heis-ham}
\hat{{\cal H}} = J \sum_{\langle ij \rangle} 
{\bf \hat{S}}_{i} \cdot {\bf \hat{S}}_{j},
\end{equation}
where $\langle ij \rangle$ denote sums over NN sites and ${\bf \hat{S}}_{i}$
is the spin-$1/2$ operator at site $i$. All energies will be given in units 
of $J$. 

The variational wave functions are defined by projecting noncorrelated 
fermionic states:
\begin{equation}
\label{eqn:var-wf}
|\Psi_{{\rm VMC}}(\chi_{ij},\Delta_{ij},\mu,\zeta)\rangle=
{\cal P}_{G}|\Psi_{{\rm MF}}(\chi_{ij},\Delta_{ij},\mu,\zeta)\rangle,
\end{equation}
where ${\cal P}_{G}=\prod_{i}(1-n_{i,\uparrow}n_{i,\downarrow})$ is the full
Gutzwiller projector enforcing the one fermion per site constraint.
Here, $|\Psi_{{\rm MF}}(\chi_{ij},\Delta_{ij},\mu,\zeta)\rangle$ is the ground
state of mean-field Hamiltonian containing chemical potential, hopping, and 
{\it singlet} pairing terms:
\begin{eqnarray}
\label{eqn:MF0}
{\cal H}_{{\rm MF}} &=
&\sum_{i,j,\alpha} (\chi_{ij}+\mu\delta_{ij})c_{i,\alpha}^{\dagger}c_{j,\alpha}\nonumber \\
&+&\sum_{i,j} \{(\Delta_{ij}+\zeta\delta_{ij})c^{\dagger}_{i,\uparrow}c^{\dagger}_{j,\downarrow}+h.c.\} \, ,
\end{eqnarray}
where $\chi_{ij}=\chi_{ji}^{*}$ and $\Delta_{ij}=\Delta_{ji}$. Besides the 
chemical potential $\mu$, we will also consider real and imaginary
components of on-site pairing, which are absorbed in $\zeta$.
We briefly mention that a somewhat similar approach, based upon a bosonic
representation of the spin operators (i.e., through Schwinger bosons), has been
also used recently.~\cite{Motrunich-2011} In the latter case, however, the
bosonic nature of quasi-particle operators implies that one has to deal with
permanents instead of determinants, which makes the numerical calculations
much heavier than in our fermionic case.

\begin{figure}
\includegraphics[width=0.95\columnwidth]{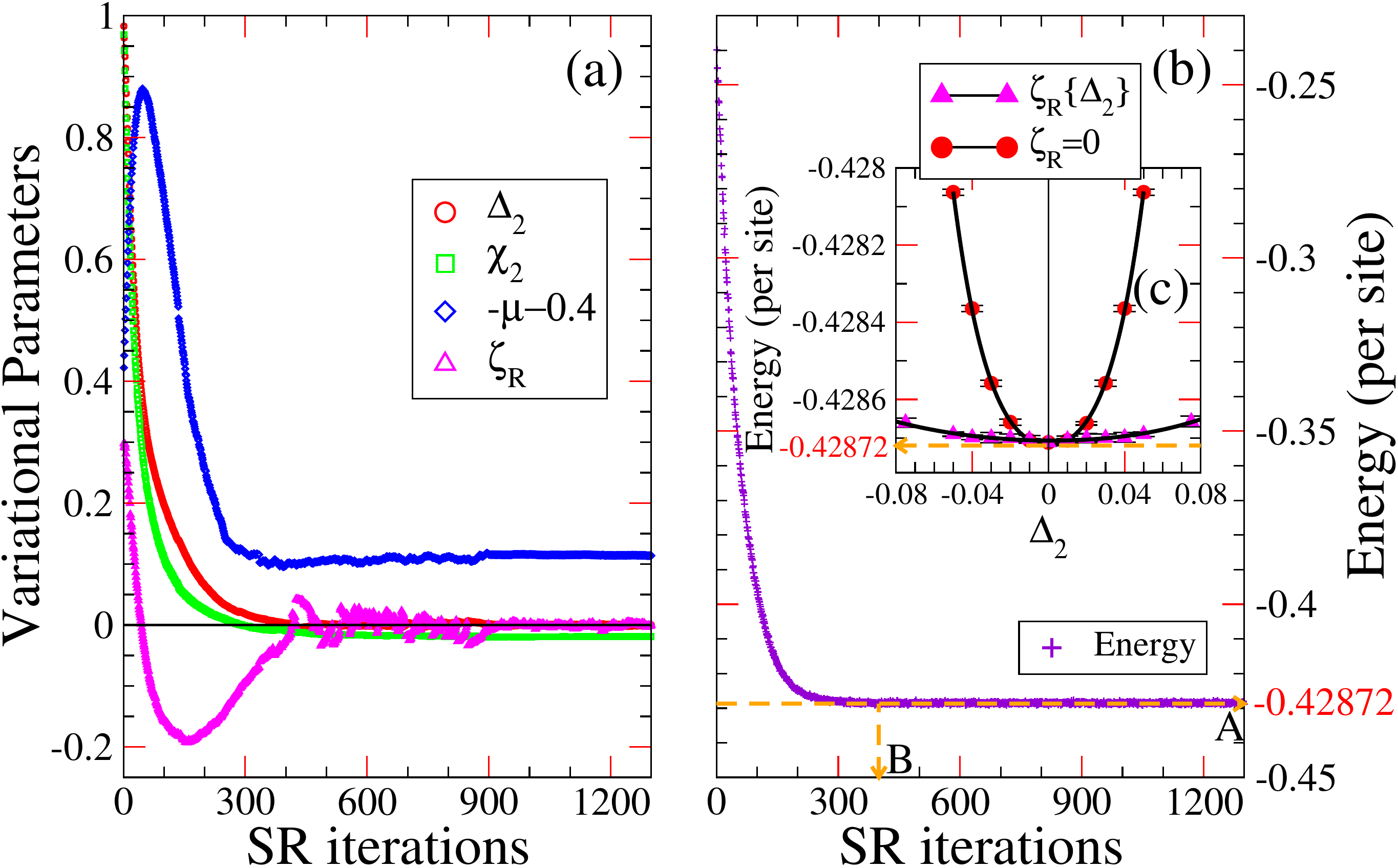}
\caption{(Color online) A typical variational Monte Carlo optimization run for
the $\mathbb{Z}_{2}[0,\pi]\beta$ wave function:~(a) variational parameters 
$\Delta_{2}$, $\chi_{2}$, $\mu$, and $\zeta_{{\rm R}}$ and (b) energy, as a 
function of SR iterations. In (a), the initialized parameter values are: 
$\Delta_{2}=\chi_{2}=1$, $\mu=-0.8$, and $\zeta_{{\rm R}}=0.3$. 
The U($1$) $2$nd NN $[0,\pi;0,\pi]$ Dirac SL corresponds to $\Delta_{2}=0$, 
$\chi_{2}=-0.0186(2)$, $\zeta_{{\rm R}}=0$, as found in 
Ref.~\protect\onlinecite{Iqbal-2011}.
The optimized parameter values are obtained by averaging over a much larger 
number of converged SR iterations than shown above. In (c), the variation
in energy upon addition of a small $\Delta_{2}$ (both for $\zeta_{\rm {R}}=0$, 
and optimized $\zeta_{\rm {R}}$ for each value of $\Delta_{2}$) upon the 
$[0,\pi;0,\pi]$ Dirac SL is shown, the increase in energy is apparent.}
\label{fig:pic2}
\end{figure}

Different SL phases correspond to different patterns of distribution of 
$\chi_{ij}$ and $\Delta_{ij}$ on the lattice links, along with the 
specification of the on-site terms $\mu$ and $\zeta$. Then, a complete 
specification of a SL state up to $n$th NN amounts to specifying the SU($2$) 
flux through closed loops along with the optimized hopping and pairing 
parameters at each geometrical distance.~\cite{Wen-2002,Wen-1991}
These parameters are the {\it Ans\"atze} of a given state and serve as the 
variational parameters in the physical wave function that are optimized 
within the variational Monte Carlo scheme to find the energetically best 
state. It is worth mentioning that we use a sophisticated implementation of 
the stochastic reconfiguration (SR) optimization 
method,~\cite{Sorella-2005,Sorella-2006} which allows us to obtain an extremely
accurate determination of variational parameters. Indeed, small energy 
differences are effectively computed by using a correlated sampling, which
makes it possible to strongly reduce statistical fluctuations.
The current problem of the study of the instability of a U($1$) Dirac SL state toward 
the $\mathbb{Z}_{2}[0,\pi]\beta$ state will clearly demonstrate the power of 
this method to capture the essential subtleties.

{\it Results}.
We performed our variational calculations on a $432$-site cluster with 
mixed periodic-antiperiodic boundary conditions which ensures nondegenerate 
wave functions at half-filling. The large size of the cluster ensures that 
the spatial modulations induced in the observables by breaking of rotational
symmetry (due to mixed boundary conditions) remain smaller than the uncertainty
in the Monte Carlo simulations.

Among the class of NN fully symmetric and gapless SLs, the U($1$) Dirac state
has the lowest energy. Its energy per site is $E/J=-0.42863(2)$, and its 
{\it Ansatz} is given by the sign convention for NN bonds in 
Fig.~\ref{fig:pic1}. Due to the U($1$) flux $\varphi$ being $0$ and $\pi$ 
[$\exp{(i\varphi)}=\prod_{{\rm plaquette}}\chi_{ij}$] through triangles and 
hexagons, respectively, it is denoted as $[0,\pi]$. Another competing state, 
the NN uniform RVB state has zero flux through any plaquette and is 
therefore denoted as $[0,0]$; its energy per site is 
$E/J=-0.41216(1)$.~\cite{Lee-2007,Hermele-2008} 

The study in Ref.~[\onlinecite{Lu-2011}] identified four $\mathbb{Z}_{2}$ SLs 
in the neighborhood of the $[0,\pi]$ state; only one of them, the 
$\mathbb{Z}_{2}[0,\pi]\beta$ state, was found to be gapped (via the $2$nd NN 
spinon pairing term). Its {\it Ansatz} up to $2$nd NN mean-field bond is 
reproduced in Fig.~\ref{fig:pic1}.~\cite{Lu-2011}
In a suitable gauge, its mean-field {\it Ansatz} is specified by five real 
parameters. These parameters are the $1$st NN real hopping ($\chi_{1}$), 
$2$nd NN real hopping ($\chi_{2}$), $2$nd NN real spinon pairing 
($\Delta_{2}$), and two onsite terms, one for the chemical potential $\mu$ and 
the other for the real on-site pairing $\zeta_{{\rm R}}$. The mean field 
Hamiltonian can be then conveniently cast in the following form:
\begin{eqnarray}
&&{\cal H}_{{\rm MF}}\{\mathbb{Z}_{2}[0,\pi]\beta\} =
\chi_{1}\sum_{\langle ij\rangle,\alpha}{\rm s}_{ij} c_{i,\alpha}^{\dagger}c_{j,\alpha} \nonumber \\&+&
\sum_{\boldsymbol\langle\langle ij\rangle\boldsymbol\rangle}\nu_{ij}\{\chi_{2}\sum_{\alpha}c^{\dagger}_{i,\alpha}c_{j,\alpha}+
\Delta_{2}(c^{\dagger}_{i,\uparrow}c^{\dagger}_{j,\downarrow}+ h.c.)\}\nonumber \\
&+&\sum_{i}\{\mu\sum_{\alpha}c_{i,\alpha}^{\dagger}c_{i,\alpha}
+\zeta_{{\rm R}} (c_{i,\uparrow}^{\dagger}c_{i,\downarrow}^{\dagger}+h.c.)\},
\label{eqn:MF-Z2}
\end{eqnarray}
where $\langle ij\rangle$ and 
$\boldsymbol\langle\langle ij\rangle\boldsymbol\rangle$ denote sums over $1$st 
and $2$nd NN sites, respectively. ${\rm s}_{ij}$ and $\nu_{ij}$ encode the 
sign structure of the $1$st and $2$nd NN bonds, respectively, as shown in 
Fig.~\ref{fig:pic1}. The $1$st NN real hopping ($\chi_{1}$) will be taken as 
a reference, and hence set to unity hereafter. The physical variational wave 
function of this SL state then depends on four variational parameters, 
$|\Psi_{{\rm VMC}}(\chi_{2},\Delta_{2},\mu,\zeta_{R})\rangle=
{\cal P}_{G}|\Psi_{{\rm MF}}(\chi_{2},\Delta_{2},\mu,\zeta_{R})\rangle$.

For a generic unbiased starting point in the four-dimensional variational 
space, the variation of parameters and energy in the SR optimization is shown 
in Fig.~\ref{fig:pic2}. As one can clearly see, the energy converges neatly 
[see point B in Fig.~\ref{fig:pic2}(b)] to the reference value of the 
suitably extended $2$nd NN U($1$) Dirac SL, the $[0,\pi;0,\pi]$ state 
[see point A in Fig.~\ref{fig:pic2}(b)], with small but finite $\chi_{2}$ 
[see Fig.~\ref{fig:pic2}(a)] previously computed by us.~\cite{Iqbal-2011} 
For the present cluster, these values are $E/J=-0.42872(1)$ per site, and 
$\chi_{2}=-0.0186(2)$, $\mu=-0.5124(5)$. Also, it is manifest that 
$(\Delta_{2}$, $\zeta_{{\rm R}})\rightarrow0$, becoming exactly zero 
(within the error bars) after averaging over a sufficient number of converged 
Monte Carlo steps. Here, we bring attention to the important fact that, 
despite the energy having converged after $\approx 400$ iterations, the 
parameters did not converge and were still varying, converging to their final values 
much later than the energy (see Fig.~\ref{fig:pic2}). This fact is possible 
because, in the energy minimization, forces are calculated through the 
correlated sampling and not by energy differences.~\cite{Sorella-2005} 
Our result shows that the energy landscape along the manifold connecting the 
U($1$) Dirac SL to the $\mathbb{Z}_{2}[0,\pi]\beta$ SL is {\it very} flat 
close to the U($1$) Dirac SL [see Fig.~\ref{fig:pic2}(c) for the case 
$\zeta_{{\rm R}}\{\Delta_{2}\}$]. Consequently, a small perturbation around 
the U($1$) Dirac SL, e.g., by setting $\Delta_{2}=0.05$ along with the 
corresponding optimized value of $\zeta_{{\rm R}}=0.1780(2)$ will not lead 
to any detectable change in energy. Hence, one cannot unambiguously conclude 
anything about the stability of the U($1$) Dirac SL by solely computing the 
energy of the perturbed wave function with fixed parameters, point by point 
locally. Only by performing an accurate SR optimization 
method~\cite{Sorella-2005} can one successfully optimize the parameters and 
transparently show that $\Delta_{2}=0$ corresponds to the actual minimum of 
the variational energy.
This fact implies that the U($1$) gauge structure is kept intact and the Dirac
SL state is {\it locally} and {\it globally} stable with respect
to destabilizing into the $\mathbb{Z}_{2}[0,\pi]\beta$ state. We verified this
result by doing many optimization runs starting from different random initial
values of the parameters in the four-dimensional variational space.~\cite{Iqbal-comment}

Regarding the remaining three gapless $\mathbb{Z}_{2}$ SLs in the neighborhood 
of the U($1$) Dirac SL, namely, the $\mathbb{Z}_{2}[0,\pi]\alpha$, 
$\mathbb{Z}_{2}[0,\pi]\gamma$, and $\mathbb{Z}_{2}[0,\pi]\delta$ states of 
Ref.~\onlinecite{Lu-2011}, our study reveals the same result as for the 
$\mathbb{Z}_{2}[0,\pi]\beta$ state. That is, all three of these SLs return 
back exactly to the gapless U($1$) Dirac SL state, with the value of the 
parameter responsible for breaking the U($1$) gauge structure down to 
$\mathbb{Z}_{2}$ exactly vanishing upon optimization. Thus, we can convincingly
conclude that the $\mathbb{Z}_{2}$ neighborhood of the U($1$) Dirac state 
does not contain the presumed fully gapped $\mathbb{Z}_{2}$ SL found by 
the DMRG study.

This conclusion forces us to shift our focus to the $\mathbb{Z}_{2}$ 
neighborhood of another fully symmetric (and energetically competing) gapless 
SL, called the uniform RVB or the $[0,0]$ SL. Despite having a slightly higher
energy, it has the promising feature that all four $\mathbb{Z}_{2}$ SLs in 
its neighborhood are gapped, thereby opening up the possibility, albeit a slim 
one, that opening a gap might lead to a large gain in energy so as to make 
one of these four states go lower than the U($1$) Dirac SL, near to the DMRG 
value of $E/J=-0.4379(3)$ per site. These gapped SLs are referred to in 
Ref.~[\onlinecite{Lu-2011}] as the $\mathbb{Z}_{2}[0,0]{\rm A}$, 
$\mathbb{Z}_{2}[0,0]{\rm B}$, $\mathbb{Z}_{2}[0,0]{\rm C}$, and 
$\mathbb{Z}_{2}[0,0]{\rm D}$ states; for their {\it Ans\"atze}, see Table I in
Ref.~[\onlinecite{Lu-2011}] and also the supplementary material.
Our simulations show that all four of these SLs return upon optimization to 
the gapless uniform RVB SL, with optimized $\chi_{{\rm n}}$. In particular, 
case by case we see that: for the $\mathbb{Z}_{2}[0,0]{\rm A}$ SL, the $2$nd NN
spinon pairing term goes to zero along with the on-site pairing term,
thus returning back to the $2$nd NN uniform RVB SL, the $[0,0;\pi,\pi]$ state 
given by optimized $\chi_{2}=-0.032(1)$;~\cite{Iqbal-2011} 
for the $\mathbb{Z}_{2}[0,0]{\rm B}$ SL, the NN spinon pairing term goes to 
zero upon optimization, thus giving back the NN uniform RVB SL; the 
$\mathbb{Z}_{2}[0,0]{\rm C}$ SL upon optimizing flows to the $3$rd NN uniform 
RVB SL with optimized $\chi_{2}$ and $\chi_{3}'$s, with the spinon pairing term 
at $3$rd NN becoming zero; and finally, the $\mathbb{Z}_{2}[0,0]{\rm D}$ SL flows 
back to the $2$nd NN uniform RVB SL. The results showing how the energies of 
these extended gapless uniform RVB SLs increase as the 
${\rm U}(1)\rightarrow\mathbb{Z}_{2}$ gauge breaking parameter is tuned on 
from zero to a small finite value are reported in the supplementary material 
(at the end of the article).

For reasons of completeness, we mention that there are two more 
gapless U($1$) SLs in whose neighborhoods the remaining four $\mathbb{Z}_{2}$ SLs 
(all gapless) lie.~\cite{Lu-2011} However, these U($1$) SLs suffer from a 
{\it macroscopic} degeneracy at half-filling which leads to an open shell. 
This degeneracy being macroscopic cannot be removed by using any of the four 
real boundary condition possibilities. Hence, their energy can only be computed
approximately in the limit by inserting an additional $\theta$ flux through 
the triangle motifs and consequently removing $\pi-2\theta$ through hexagon 
motifs, and then taking the limit $\theta\rightarrow0$. The energy of the 
SL-$[\pi,\pi]$ computed in this way is $E/J \simeq -0.38372(1)$ per site, 
which is much higher than those of other gapless U($1$) states. Hence, we did 
not carry out an analysis of $\mathbb{Z}_{2}$ SLs in these two neighborhoods.

Finally, we also investigated the possibility of stabilization of the 
$\mathbb{Z}_{2}[0,\pi]\beta$ state upon addition of a small $2$nd NN 
exchange coupling ($J^\prime$) of both antiferromagnetic and ferromagnetic 
type in the NN spin-$1/2$ QHAF. In both cases, on optimization we found that
$(\Delta_{2}$, $\zeta_{{\rm R}})\rightarrow0$, becoming exactly zero 
(within error bars) after averaging over a sufficient number of converged
Monte Carlo iterations. Thus, we recover the suitably extended, gapless 
$2$nd NN U($1$) Dirac SL. In particular, for $J^\prime/J=0.10$, this is the
$[0,\pi;\pi,0]$ state with optimized $\chi_{2}=0.0924(2)$, and 
$E/J=-0.43200(2)$ per site; for $J^\prime/J=-0.10$, this is the 
$[0,\pi;0,\pi]$ state with optimized $\chi_{2}=-0.1066(2)$, and 
$E/J=-0.42898(2)$.~\cite{Iqbal-2011}    
 
In summary, we investigated the possibility of stabilizing gapped 
$\mathbb{Z}_{2}$ SLs in the NN and next-nearest-neighbor (NNN) spin-$1/2$ QHAF on a kagome lattice.
We found that none of the five gapped $\mathbb{Z}_{2}$ SLs [one connected to
the U($1$) Dirac state and the other four connected to the uniform RVB state]
can occur as ground states. In particular, the most promising gapped SL 
conjectured to describe the ground state, the $\mathbb{Z}_{2}[0,\pi]\beta$ 
state, is always higher in energy than the U($1$) Dirac SL. 
Our systematic numerical results bring us to the conclusion that, at least
within the Schwinger fermion approach of the spin model, the U($1$) Dirac SL
has the best variational energy for the NN and NNN spin-$1/2$ QHAF on 
kagome lattice. The conflict of our results, which point toward a gapless 
ground state, and the ones by recent DMRG calculations, which instead suggested
the presence of a fully gapped spectrum, remains open and deserves further
investigations. One possible direction would be to consider further 
improvements of our variational wave functions, based upon the application of
few Lanczos steps or an approximated (fixed-node) projection technique, which,
e.g., gives an energy of $E/J=-0.431453(2)$ per site for the NN U($1$) Dirac
SL, and $E/J=-0.431443(2)$ for the NNN $[0,\pi;0,\pi]$ state. Another 
possible direction would be to explore the energetics of gapped 
$\mathbb{Z}_{2}$ SLs which break some symmetries such as time-reversal. 
The possibility that the fully gapped SL found by the DMRG study possesses a 
different low energy gauge structure other than $\mathbb{Z}_{2}$ also remains
open.



\section*{Supplementary material}

# Supplementary material for
# "Projected wave function study of $\mathbb{Z}_2$ spin liquids on the kagome lattice for the spin-$\frac{1}{2}$ quantum Heisenberg antiferromagnet"

Yasir Iqbal,[1] Federico Becca,[2] and Didier Poilblanc[1]

[1]*Laboratoire de Physique Théorique UMR-5152, CNRS and Université de Toulouse, F-31062 France*
[2]*Democritos National Simulation Center, Istituto Officina dei Materiali del CNR and Scuola Internazionale Superiore di Studi Avanzati (SISSA), Via Bonomea 265, I-34136 Trieste, Italy*

(Dated: July 21, 2011)

In this supplementary material, we present the explicit plots showing the variation of energy as a function of the $U(1) \to \mathbb{Z}_2$ gauge breaking parameters for the four gapped $\mathbb{Z}_2$ SLs in the neighborhood of the gapless $[0,0]$ SL (the uniform RVB state). We also reproduce the *ansatz* (from Table I of Ref. 21) of the five SLs investigated in paper, so as to make the paper self contained.

| State | $\Lambda_{\text{onsite}}$ | $U_{\text{n.n.}}$ | $U_{2nd\text{n.n.}}$ | $U_{3rd\text{n.n.}}$ | $\tilde{U}_{3rd\text{n.n.}}$ |
|---|---|---|---|---|---|
| $\mathbb{Z}_2[0,\pi]\beta$ | $\mu, \zeta_R$ | $\chi_R$ | $\chi_R, \Delta_R$ | 0 | 0 |
| $\mathbb{Z}_2[0,0]A$ | $\mu, \zeta_R$ | $\chi_R$ | $\chi_R, \Delta_R$ | 0 | 0 |
| $\mathbb{Z}_2[0,0]B$ | $\mu$ | $\chi_R, \Delta_I$ | 0 | 0 | 0 |
| $\mathbb{Z}_2[0,0]C$ | $\mu$ | $\chi_R$ | $\chi_R$ | $\chi_R, \Delta_I$ | $\chi_R$ |
| $\mathbb{Z}_2[0,0]D$ | $\mu$ | $\chi_R$ | $\chi_R, \Delta_I$ | 0 | 0 |

TABLE I. The mean field *ansatz* of the five gapped SLs investigated by us, given only up to the neighbor at which the gauge symmetry is broken, in a form used by us in numerical simulations. The parameters highlighted in red are responsible for opening a gap by breaking the $U(1)$ gauge symmetry down to $\mathbb{Z}_2$. The $U_{3rd\text{n.n.}}$ denotes bonds of length 2 connecting two sites and passing through a third site (such as the bond $1 \to 4$ in Fig. 1 in the main paper); instead, $\tilde{U}_{3rd\text{n.n.}}$ denotes bonds of length 2 which dont pass through any site.

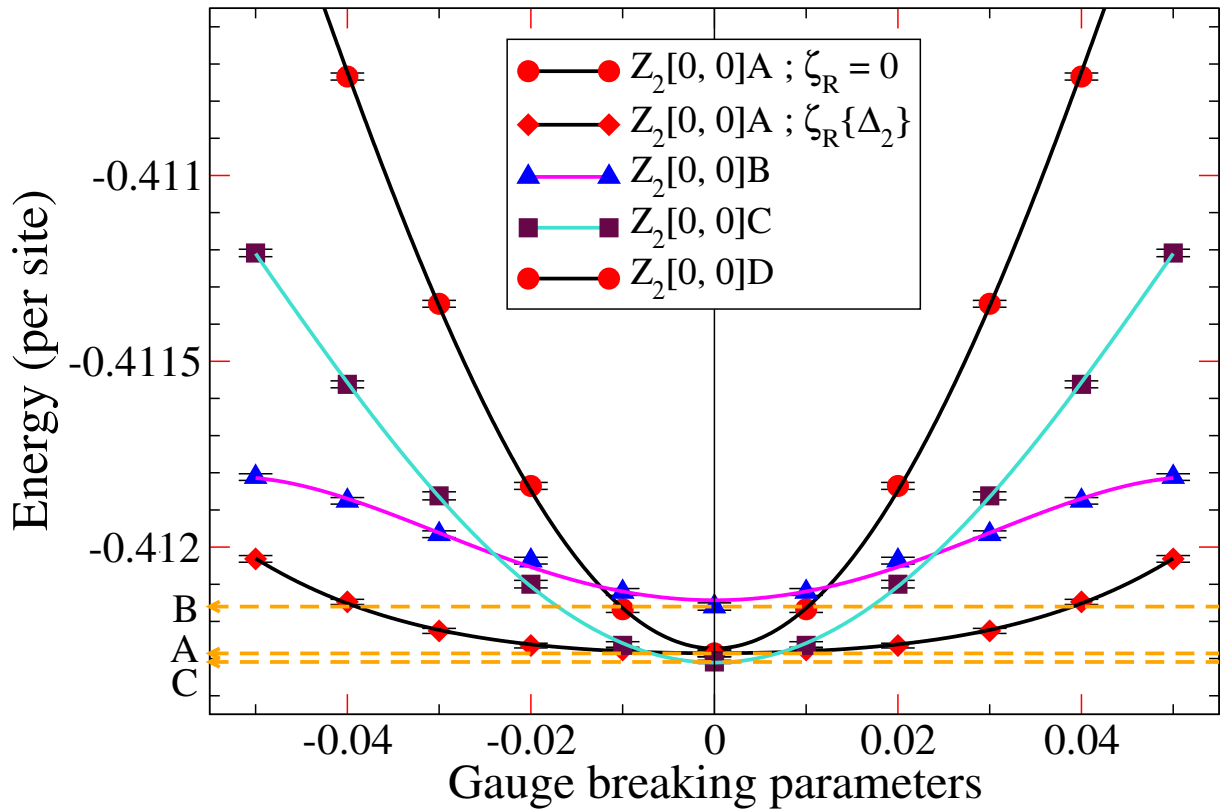

FIG. 1. For the four $\mathbb{Z}_2$ spin liquids in the neighborhood of the $[0,0]$ state: The manner of variation of energy as the gauge breaking parameter (shown in red in Table I) is tuned on from zero to a small finite value, is shown. The parameters in black in Table I are fixed to their optimized values, and correspond to a suitably extended $n^{\text{th}}$ NN gapless uniform RVB SL. The increase in energy upon opening a gap is apparent.

In the main paper, we have discussed the numerical results of a full Monte Carlo optimization for the four gapped $\mathbb{Z}_2$ SLs in the neighborhood of the $[0,0]$ state. In particular, we find that, by starting from arbitrary points in the variational space, one returns back to the gapless $[0,0]$ reference SL (suitably extended to $n^{\text{th}}$ NN), with the value of the gauge breaking parameter going to zero. The energies of the NN $[0,0]$ SL and the extended 2nd and 3rd n.n. $[0,0]$ SLs correspond to the points B, A, C in Fig. 1 respectively, these energies are $E/J = -0.41216(1)$, $E/J = -0.41228(1)$, $E/J = -0.412308(1)$ respectively. Here, we show the *local* explicit manner of increase in energy of these extended gapless $[0,0]$ SLs upon addition of a *small* gap opening (gauge breaking) parameter (given in red in Table. I), like we did for the $\mathbb{Z}_2[0,\pi]\beta$ state in Fig. 2(c) of main paper. It should be noted that the case of $\mathbb{Z}_2[0,0]A$ with $\zeta_R = 0$ is equivalent to the $\mathbb{Z}_2[0,0]D$ case. Hence, we have the same plot for both SLs in Fig. 1.



\end{document}